# Electronic structure of Fe (0-5 at. %) doped $MoO_2$ thin films studied by resonant photoemission spectroscopy

Ram Prakash, R. J Choudhary and D.M. Phase[*]

UGC-DAE Consortium for Scientific Research,

University Campus, Khandwa Road, Indore-452001, India.

**Abstract**

The electronic structure of pulsed laser deposited $Mo_{1-x}Fe_xO_2$ ($x$=0, 0.02 and 0.05) thin films has been investigated using resonant photoemission spectroscopy near Mo-4p absorption edge. In all the samples a broad Fano- like resonance peak at ~46 eV is observed in the whole area of the valence band which indicates the contribution of the Mo-4d states in the entire valence band region. The doping of Fe in these films leads to decrease in Mo 4d states contributing to electronic states at lower binding energy region. In addition to this, we also observe a shoulder at 4.9 eV in the valence band spectra of doped samples. It is proposed that the origin of this shoulder is due to the Fe hybridised states.

---

[*] Correspondence author e mail : dmphase@csr.ernet.in

**Introduction**

The transition metal oxides exhibit a variety of interesting physical properties [1]. One particular interesting phenomena is that these oxides show a wide range of electrical conductivity, which varies from insulators (e.g.,NiO,$TiO_2$, $MoO_3$) to semiconductor (e.g.,$NbO_2$) to conductors (e.g. $WO_2$, $MoO_2$ ). Among these compounds molybdenum oxide shows variable electrical conductivity from metallic $MoO_2$ to insulating $MoO_3$ phase with variation in oxygen content [2-3]. Recently molybdenum oxide has attracted much interest because of its unique properties and applications in gas sensing devices [4,5], optically switchable coatings[6], catalysis[7] etc. In $MoO_2$, another degree of dimensionality in terms of magnetic property may be induced by doping some magnetic impurity. The resulting device material will be the new dilute magnetic oxide for $MoO_2$ based present technologies. In our recent paper [8] we reported the room temperature ferromagnetism in (100) oriented Fe doped molybdenum oxide thin films grown on c-axis sapphire substrate using pulsed laser deposition. The x-ray photoelectron spectroscopy revealed that Fe is in +2 state, which favours the substitutional occupancy of Fe ion in the $MoO_2$ matrix.

From the fundamental physics point of view as well as to find out the origin of ferromagnetism in iron doped samples, it is essential to have a knowledge of electronic structure of both undoped and doped samples particularly around the Fermi level. The electronic structure of $MoO_2$ has been investigated theoretically by means of phenomenological molecular orbital scheme as well as by tight-binding and cluster calculations [9-12]. Experimentally it is shown by various spectroscopic measurements that 9 eV wide occupied band can be divided into two energy regions, 3 eV wide Mo 4d bands at low binding energy region and 6 eV wide O 2p bands at higher binding energies [2,10,12-15]. Resonant photoemission spectroscopy (RPES) is a well known technique to obtain the detailed information on the valence electronic structure of a material [16-19]. To best of our knowledge, there is hardly any report available on RPES measurements on doped or un-doped $MoO_2$ in existing literature. In the present paper, we report our results of resonance photoemission measurements on pulsed laser deposited Fe (0-5 at %) doped $MoO_2$ thin films. The energy of incident photon is varied around the Mo 4p→ 4d threshold. The contribution of Mo 4d states to the various regions of the valence band and effect of Fe doping is discussed on the basis of the analysis of the resonant behaviour of the valence band emission.

**Experimental**

Thin films of undoped and Fe (2 and 5 at %) doped $MoO_2$ were grown on c-plane sapphire single crystal substrates by pulsed laser ablation technique as dicusussed in reference [8]. X-ray diffraction (XRD) revealed that all the samples are in single phase. The homogeneity of the samples was checked with energy dispersive analysis of X-rays (EDAX) technique. The thicknesses of all the films are around 200 nm as measured by stylus profilometer. The X-ray diffraction (XRD), X-ray photoelectron spectroscopy (XPS), resistivity and magnetization measurements were performed for characterization of these samples. These results are reported elsewhere [8]. The RPES measurements on these samples were carried out at angle integrated PES beamline on Indus-1 synchrotron radiation source. The valence band photoemission spectra were recorded using photon energy of 34 to 68 eV in the step of 2 eV. All the spectra were measured using an Omicron (EA125) energy analyser. The angle between incoming photon and emitting electron was 90°, while the angle between sample and photon beam was 45°. The acceptance angle of spectrometer was $\pm 8°$ during the measurements. All measurements were carried out at room temperature. The spectra presented below were normalized by photon flux estimated from the photocurrent from post mirror of the beam line. The background pressure in the analysis chamber was $1\times10^{-10}$ mbar. The sample surface was cleaned using 1 keV Ar ions. Before performing RPES measurements the surface composition was estimated from core level XPS measurements. The Fermi level was aligned by recording the valence band spectra of in-situ cleaned Au foil. The total experimental resolution was estimated from the Fermi edge of gold foil. The experimental resolution, dependent on the photon energy, was estimated to 0.32 to 0.44 eV in the photon energy range 34 to 68 eV.

**Results and discussion**

Figure1 shows the valence band (VB) spectra of undoped and Fe (2 and 5 at. %) doped $MoO_2$ thin films, recorded with 34 eV photon energy. The background obtained by the Shirley procedure [20] has been subtracted from each raw data. All the spectra show finite density of states at Fermi level indicating metallic nature in confirmation with the resistivity measurements [8] of these films. The width of the VB spectra is close to the reported theoretical and experimental results [12,13]. The VB spectra corresponding to all the three films consist of two dominant bands and a hump; the lower binding energy band peaked at 1.8eV and the higher binding energy band peaked at 5.5 eV and hump at 6-8 eV. These spectra match well with the reported spectra of $MoO_2$ (Bulk) [10,13]. Comparison of the observed spectra with that of theoretically calculated density of states (DOS) [11-12] indicates that lower binding energy band is mainly derived from Mo 4d states while higher energy band is mostly O 2p states derived. A hump at 6-8 eV is derived from hybridization of O 2p and Mo 4d states. On comparing the valence band spectra of undoped and Fe doped sample a small shoulder at 4.9eV binding energy was observed. Normally Fe-3d states are observed around 1-3 eV, but since 0-3 eV region is dominated by Mo 4d states and Fe concentration is very small, therefore, these sates are not visible in the valence band spectra. In order to understand the electronic structure of $MoO_2$ thin film and effect of Fe doping in detail we have carried out RPES measurements on both undoped and doped $MoO_2$ films by varying the incident photon energy in the Mo 4p→4d absorption region.

Figures 2 (a, b and c) show valence band spectra of undoped, 2% Fe doped, and 5% Fe doped $MoO_2$ thin films taken at various photon energies respectively. In order to further analyse the data we divide the valence band spectra in four regions viz (i) near the fermi edge (0.15 eV) feature A, (ii) the peak region of lower binding energy band (1.8eV)- feature B, (iii) the peak region of higher binding energy band (5.5 eV)-feature C and (iv) the hump region (7.1 eV) – feature D. Each spectrum in the Fig. 2 is normalized by its peak intensity of feature C. We have normalized the spectra with intensity of feature C because this feature is mainly due to O 2p derived states and resonance is recorded at Mo 4p→4d absorption edge. Further photoemission cross section for O 2p state does not change appreciably in such narrow energy range [21]. Thus it is expected that intensity of feature C would not change significantly. It is evident from the figure when the photon energy is varied from 34 to 38 eV, a marginal decrease in the intensity of features A, B and D is noticeable. The intensity of these features increases monotonically for photon energy between 42 and 48 eV. The intensity of these features again decreases with further increase in photon energy. Further, increase in the intensity of features A and B is much more significant in comparison to the increase in the intensity of feature D. Similar variations in the VB spectra with photon energy have been observed in all the 2% and 5% Fe doped $MoO_2$ films. (Figs. 2b and 2c)

To evaluate the resonance energy and effect of iron doping we have plotted constant initial state (CIS ) intensity plots for features A, B and D for all the three films in figure 3. The CIS intensity plots shown in figure 3 and 4 are obtained from the figure 2 by plotting the normalised intensity of marked regions (A, B and D) at fixed binding energy positions of respective features. A maximum in the intensity of these spectra around photon energy 46 eV is observed for features A, B and D. This resonance energy matches with the previously reported resonance value for Mo 4p→ 4d absorption region for $Mo_2C$ single crystal [22-24]. The observed hν-dependence is the characteristic of the Fano line shape [25], and maximum of the photoemission intensity is interpreted as originating from the interference between the normal Mo 4d photoemission process and the process induced by the photon induced excitation, $Mo4p^6 4d^n$ + hν →$Mo4p^5 4d^{n+1}$ , followed by the emission of a Mo 4d electron through a super-Coster-Kronig decay, $Mo4p^5 4d^{n+1}$ → $Mo4p^6 4d^{n-1}$ + $e^-$ . In a similar resonant photoemission study of α-$Mo_2C$ (0001) performed by Sugihara et al.[23], they found a resonance peak at 46 eV due to Mo 4p→4d transition and observed the contribution of Mo 4d orbital included in the whole area of valence band region. In another study Lince et al. [26] performed a resonant photoemission study for $MoS_2$ and found that the Mo-4d derived non-bonding band shows a resonance due to Mo 4p→4d transition at ~ 42 eV. Our results are in agreement with these studies [22-24].

The energy dependence of the feature A is shown in Fig 3 (a) for all the three samples. For Fe doped samples the behaviour of feature A is similar to the undoped $MoO_2$ film showing resonance at 46 eV. Moreover, with the increase in Fe content a decrease in the intensity of feature A is evident from the figure. Variation in the intensity of feature B with photon energy is shown in Fig. 3 (b) for all samples. The resonance effect here also is similar to that of feature A. Like in case of feature A, in feature B also a decrease in the intensity is observed with Fe doping while over all behaviour of the feature with photon energy remains same. In Fig. 3(c) we show the variation in intensity of feature D with incident photon energy. The feature D also has a resonance peak at ~ 46 eV similar to features A and B but resonance effect is weaker in comparison to the one observed for features A and B. Moreover, in contrast to the behaviour of intensity of features A and B, it can be noted here that the intensity of feature D increases with Fe doping.

From the above observations it is evident that all the three features A, B and D in the $MoO_2$ exhibit resonance around 46 eV. The resonance is stronger for features A and B in comparison to feature D. This is in confirmation with the earlier reported band structure calculations according to which features A and B are mainly due to Mo 4d states while both Mo and O 2p hybridized states contribute to the feature D. When Fe is doped into $MoO_2$ films, the effect of doping is clearly reflected in these CIS spectra. With the increase in Fe content the resonance effect for features A and B weakens while for the feature D an enhancement in the resonance effect is observed. This would indicate depletion in Mo 4d states contribution to features A and B. These states are possibly partly replaced by Fe 3d states, which may be the reason for the increase in the conductivity and observation of ferromagnetism in the doped films. The increase in intensity of feature D with Fe doping suggests an enhanced hybridization of Mo 4d and O2p states. In our XPS study on these films [8] it is observed that after Fe doping, $Mo^{4+}$ contribution increases. Since the ionic radius of $Mo^{4+}$ is larger than $Mo^{5+}$ or $Mo^{6+}$, we may expect an enhanced hybridization. However, to further verify this claim, theoretical band structure calculations on Fe doped $MoO_2$ are needed.

To understand the origin of the shoulder at 4.9 eV (feature E in Figs. 2b and 2c), we have plotted CIS intensity at binding energy 4.9 eV versus the incident photon energy for all the samples (Fig. 4). These CIS intensity plots show an increase in the intensity at around 56 eV for Fe doped samples, which happens to be the Fe resonance energy (Fe3p→3d absorption region). This resonance is absent in the undoped $MoO_2$ films. These observations suggest that the feature at 4.9 eV is an indication of Fe in hybridized state. Though, in the present work we have studied the effect of Fe doping in the electronic structure by resonant photo emission spectroscopy, to have an insight of the origin of ferromagnetism in the Fe doped samples, technique like spin-resolved photoelectron spectroscopy would be more useful.

**Conclusions**

In conclusion we have studied the electronic structure of pulsed laser deposited Fe (0-5 at %) doped $MoO_2$ thin films using resonance photoemission spectroscopy. The observed VB spectra are similar to that observed for bulk $MoO_2$. The observation of finite density of states at Fermi level of these VB spectra indicates metallic nature of the films. The Fe doping in the $MoO_2$ up to 5 at% does not influence much on valence band spectra of these films but shows a small reduction of Mo states at lower binding energy region. When the photon energy is varied from 34 -68 eV, in the Mo 4p→ 4d transition threshold region, the resonance is observed at photon energy 46 eV for all the films. We have also observed a shoulder in VB spectra of doped samples at binding energy 4.9 eV, which is attributed to Fe hybridized state.

**Acknowledgements**

Mr. A. Wadikar for help in measurements. One of us (RP) is thankful to CSIR, New Delhi, India for financial support as senior research fellowship.

**Figure captions**

**Figure 1**: Valence band spectra recorded at 34 eV photon energy of undoped and Fe doped $MoO_2$/ $Al_2O_3$ thin films grown by pulsed laser deposition technique. The inset shows the expanded view of lower binding energy region.

**Figure 2**: Valence band spectra of (a) $MoO_2$/ $Al_2O_3$, (b) $Mo_{0.98}Fe_{0.02}O_2$/ $Al_2O_3$ and (c) $Mo_{0.95}Fe_{0.05}O_2$/ $Al_2O_3$ as a function of photon energy from 34 to 68 eV. All spectra are normalized to the intensity of mainly O 2p derived feature C.

**Figure 3**: CIS photoemission intensities of (a) feature A (at binging energy 0.15 eV), (b) feature B (at binging energy 1.8 eV) and (c) feature D (at binging energy 7.1 eV) as a function of photon energy for $MoO_2$/ $Al_2O_3$, $Mo_{0.98}Fe_{0.02}O_2$/ $Al_2O_3$ and $Mo_{0.95}Fe_{0.05}O_2$/ $Al_2O_3$ respectively.

**Figure 4**: CIS photoemission intensities of feature E (at binging energy 4.9 eV) as a function of photon energy for $MoO_2$/ $Al_2O_3$, $Mo_{0.98}Fe_{0.02}O_2$/ $Al_2O_3$ and $Mo_{0.95}Fe_{0.05}O_2$/ $Al_2O_3$ respectively.

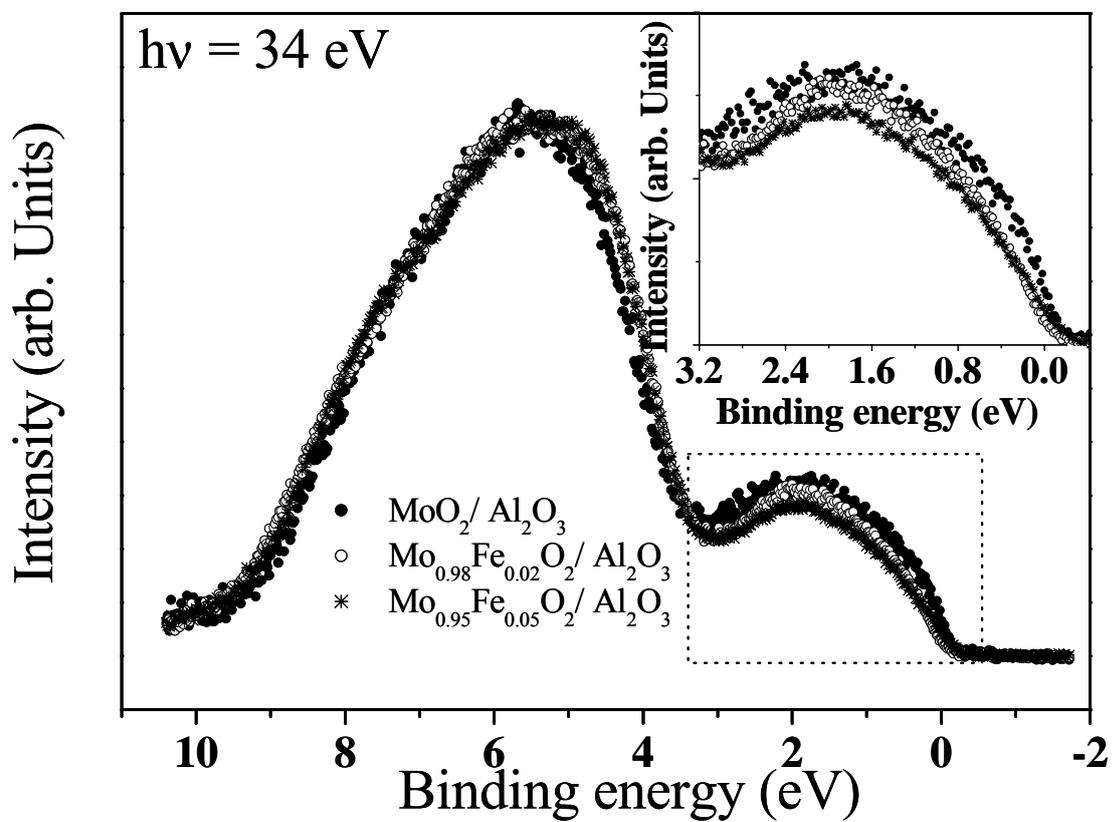

Figure 1

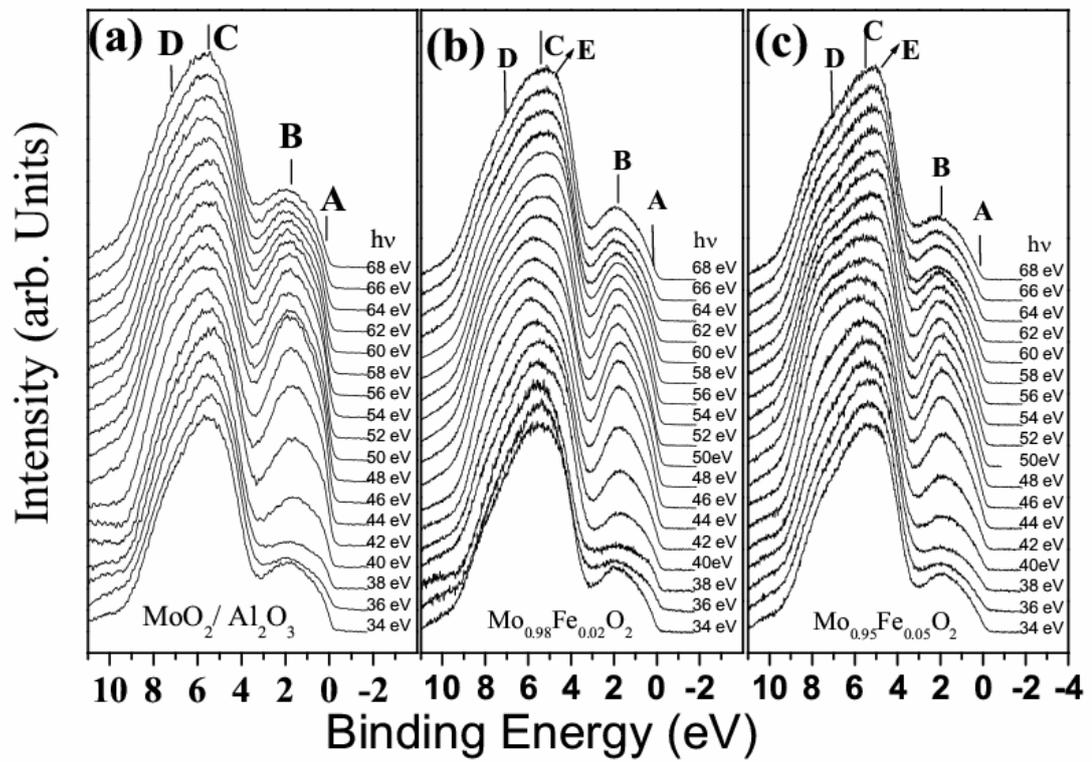

Figure 2

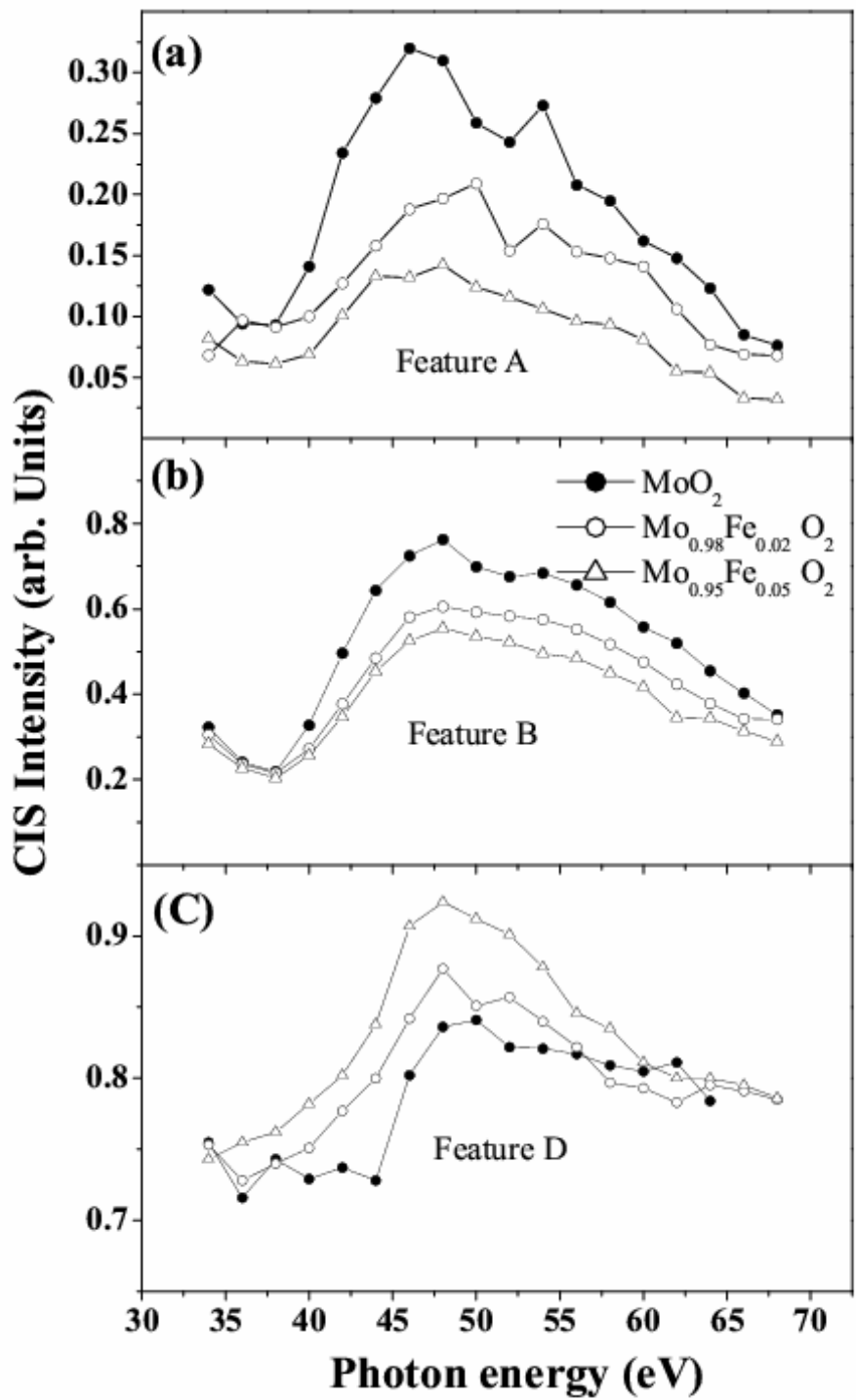

Figure 3

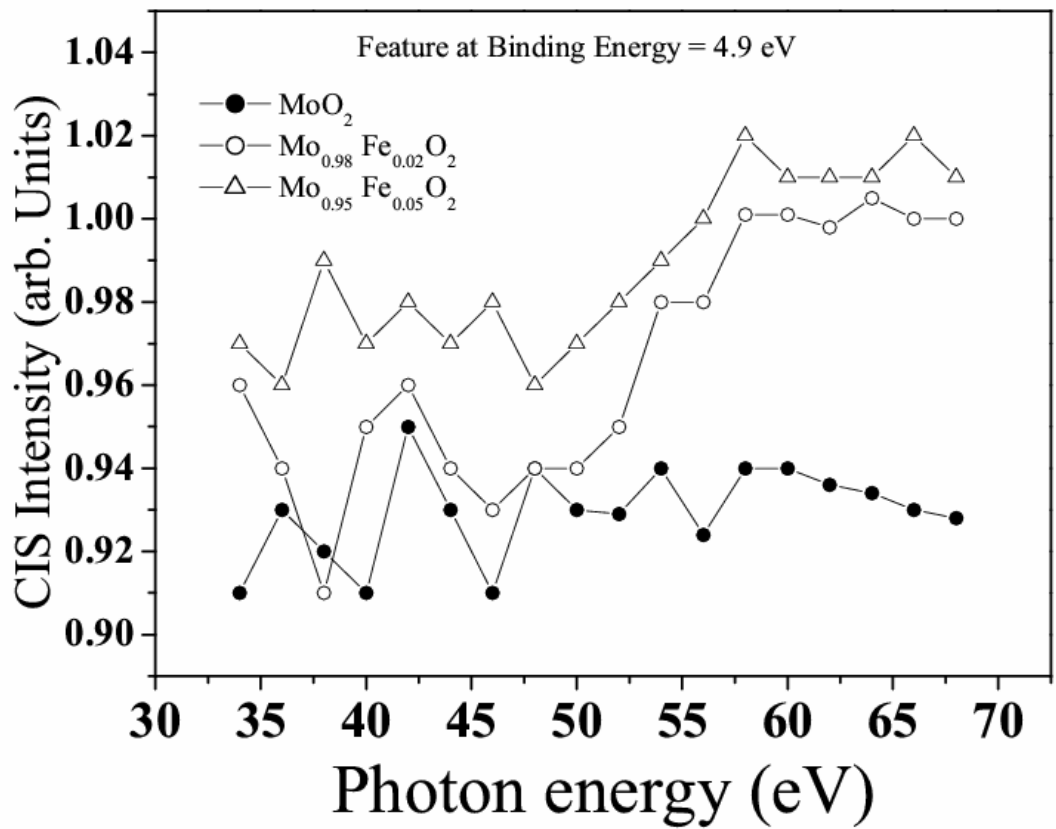

Figure 4